\documentclass[twocolumn,showpacs,aps,prl,amsmath,amssymb]{revtex4}
\usepackage{pstricks}
\usepackage{pst-coil}
\usepackage{pst-node}
\usepackage{graphicx}
\usepackage{dcolumn}
\usepackage{bm}
\usepackage{times}
\usepackage{txfonts}
\usepackage{epstopdf}
\usepackage{amsmath,epsfig}
\usepackage{float}

\newcommand{\ket}[1]{\left\vert#1\right\rangle}

\begin{document}

\title{Geometric phase kickback in a mesoscopic qubit-oscillator system }

\date{\today}

\author{G. Vacanti$^1$, R. Fazio$^{2,1}$, M. S. Kim$^3$, G. M. Palma$^4$, M. Paternostro$^5$, and V. Vedral$^{1,6,7}$ }
\affiliation{$^1$Center for Quantum Technologies, National University of Singapore, 1 Science Drive 2, Singapore\\
$^2$NEST, Scuola Normale Superiore \& Istituto di Nanoscienze-CNR, I-56126 Pisa, Italy\\
$^3$QOLS, Blackett Laboratory, Imperial College London, SW7 2BW, UK\\
$^4$NEST Istituto Nanoscienze-CNR and Dipartimento di Fisica, Univerisita' degli Studi di Palermo, Via Archirafi 36, I-90123 Palermo, Italy\\
$^5$Centre for Theoretical Atomic, Molecular, and Optical Physics, School of Mathematics and Physics, Queen's University, Belfast BT7 1NN, UK\\
$^6$Clarendon Laboratory, University of Oxford, Parks Road,  Oxford 0X1 3PU, UK \\
$^7$Department of Physics, National University of Singapore, 3 Science Drive 4, Singapore}

\begin{abstract}
We illustrate a reverse {Von Neumann measurement scheme} in which a geometric phase induced on a quantum harmonic oscillator is measured using a microscopic qubit as a probe.  We show how such a phase, generated by a cyclic evolution in the phase space of the harmonic oscillator, can be kicked back on the qubit, which plays the role of a quantum interferometer. 
We also extend our study to finite temperature dissipative Markovian dynamics and discuss potential implementations in micro and nano-mechanical devices coupled to an effective two-level system.
\end{abstract}

\maketitle

Although it is commonly believed that quantum mechanical behaviors  are preclusive characteristics of microscopic systems,  various non-classical effects  have been theoretically predicted and experimentally observed in massive systems. In opto-mechanical devices~\cite{glance,vlatko,pirandola,mauro2,mancini,marshall,vitali} as well as in setups involving superconducting qubits coupled with nano-mechanical resonators~\cite{glance2,nanooscillator}, a variety of  quantum behaviors  are observable  even at large temperature and in presence of strong dissipative processes. In general, the problem of enforcing  quantumness under unfavorable conditions  have received  great attention in the last years, being  the topic of great interest  both from a technological and from a fundamental point of view.  In this line of thoughts, we consider a system in which genuine non-classical features not only survive to temperature and dissipation, but they are indeed \emph{induced} by these environmental influences. While quantum effects like entanglement or negative values of Wigner function have been treated elsewhere~\cite{noi}, here we focus on the study of the generation and the detection of a geometric phase~\cite{berry,aharonov} in a harmonic oscillator. In the spirit of previous works~\cite{angelogp1}, we reconsider the  \emph{Von Neumann measurement scheme}, which models the measurement process as a coupling between a large measurement apparatus, used as a probe, and the  microscopic system  on which the measurement is performed, under a ``{\it reverse}'' prospective, using  the microscopic system (a qubit)  to measure the geometric phase attached to  the macroscopic one (a harmonic oscillator).

To illustrate the idea we consider a general Hamiltonian model and defer the presentation of a physical scenario suited for its realization to the second part of this work. We thus take a two-level system with logical states $\{| 0 \rangle,|1 \rangle\}$ (which we dub as the \emph{control qubit}) coupled with a harmonic oscillator  through the interaction Hamiltonian 
	\begin{equation}
\hat H = \hbar\eta |0\rangle \langle 0|\otimes (\hat{b}^\dag e^{-i \varphi} + \hat{b}e^{i \varphi}),\label{Heff}
	\end{equation}
where $b$ and $b^\dag$ are the annihilation and creation operators for the oscillator,  $\eta$ is the coupling constant  and  $\varphi$  is a phase that can be externally adjusted to change the direction of the field's quadrature in phase space. As long as $\eta$ and $\varphi$ are constant in time, the unitary operator describing the conditional time evolution of the system is 
	\begin{equation}
	\hat U_{\varphi}(t) = |1\rangle \langle 1|\otimes\hat\openone + |0\rangle \langle 0| \otimes\hat D[\zeta],\label{Ut}
	\end{equation}
where $\hat D[\zeta]=e^{\zeta\hat b^\dag-\zeta^*\hat b}$,  with $\zeta=-i\int_0^t \eta  e^{-i\varphi }$, is a displacement operator acting on the oscillator. Eq.~\eqref{Ut} describes a spin state-dependent displacement of the oscillator, which remains unperturbed if the control qubit  is in  $|1\rangle$ and is displaced by $ \zeta$ if the control qubit is in $|0\rangle$. By changing in time $\varphi$ we can drag the state of the harmonic oscillator along a nontrivial path in phase space. In particular, by varying $\varphi$ along a closed loop we can associate a purely geometric phase to the state of the harmonic oscillator. Following Ref.~\cite{wineland,milburn}, we assume that the value of $\varphi$ is changed in $n$ time steps $\delta t$ such that $t=n\delta t$, $\delta \zeta_i = {\dot \zeta_i}\delta t$, and $\hat U(t) = \prod_{i=1}^n \hat D(\delta \zeta_i )$. Recalling that $\hat D(\alpha )\hat D(\beta ) = \exp\{i\Im (\alpha\beta^* )\}\hat D(\alpha+\beta)$, we have 
\begin{equation}
\hat  U(t) = \exp\left[ i \Im\sum_{k=2}^n \delta \zeta_k \sum_{l=1}^{k-1} \delta \zeta^*_l \right]\hat D\left (\sum_{i=1}^n \delta\zeta_i \right). 
 \end{equation} 
In the continuous limit we take
$\sum_{i=1}^n \delta\zeta_i{\rightarrow}\zeta$, $\sum_{k=2}^n \delta \zeta_k \sum_{l=1}^{k-1} \delta \zeta^*_l{\rightarrow}\int \zeta^* d\zeta$
and assume that $\varphi$ is changed along a closed loop in a time $\tau$ such that 
$\zeta (\tau){=}\zeta (0)$. Making use of Stokes' theorem~\cite{nota} we find $\hat U(\tau){=}e^{i\mathcal{A}}\hat D(0) $ with $\mathcal{A}\ $ the area enclosed by the cyclic path in parameter space~\cite{notabis}.  


Let us now suppose that the system can be initialized in the state $|\psi(0)\rangle  = |+\rangle |\alpha \rangle \label{initialstate},$ where $|+\rangle = (1/\sqrt 2)(|0\rangle + |1\rangle)$ is the state of the control qubit and $|\alpha\rangle$ is a coherent state of the oscillator. We then change $\varphi (t) $ along a closed path spanned in a time $\tau$. The joint state of the qubit and oscillator at time $\tau$ will be
\begin{equation}
\label{StateUnitaryT}
|\psi(\tau)\rangle =(1/\sqrt 2)( |1 \rangle + e^{-i\mathcal{A}} |0\rangle )\otimes| \alpha \rangle.
\end{equation}

\begin{figure}[t]
{\bf (a)}\hskip3.8cm{\bf (b)}\\
\includegraphics[width=.48\textwidth]{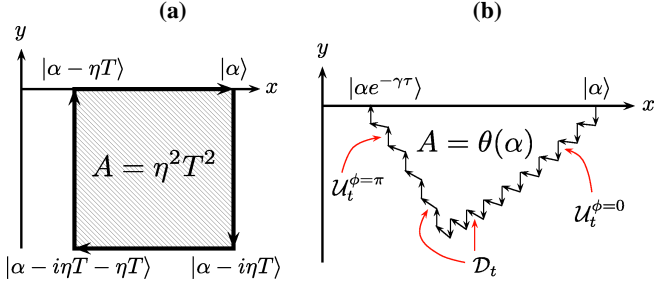}
\caption{The oscillator's conditional dynamics pictured in phase space. In {\bf (a)} the oscillator is displaced along a square whose area is proportional to the phase $\theta$. In {\bf (b)} the oscillator is displaced while undergoing a dissipative process. Here $\hat{\mathcal{U}}^{\varphi}_{\delta t}$ and $\hat{\mathcal{D}}_{\delta t}$  are the superoperators describing the unitary and dissipative evolution of duration $\delta t$, respectively.}\label{geomphase}
\end{figure}

To better understand such a result, we discuss an illustrative example where we consider a rectangular path implemented  by a stepwise change of  $\varphi$: its value is set to zero for an interval of length $T$ and then changed to $\varphi_n=n\pi/2$ at time $t = nT$ with $n=1,2,3$. As shown above, the state of the qubit acquires a phase $\vartheta = \mathcal{A} = \eta^2 T^2$ equal to the area enclosed by the path [see Fig.\ref{geomphase} {\bf (a)}]. We now stress three important points. First: although in the above example the phase explicitly depends on $T$, it is invariant with respect to the parametrization of the path. Indeed, if the area enclosed by the path remains unchanged, even for an arbitrary dependence of $\zeta ( t)$ on time we would obtain the same result $\vartheta =\mathcal{A}$.
Second: for a closed loop, $\vartheta$ does not depend on the amplitude $\alpha$ of the initial state of the harmonic oscillator. Third: the phase acquired by the harmonic oscillator is {\em kicked back} on the qubit state, which takes the role of a microscopic interferometer. As a final remark, it is important to stress that the geometric phase addressed here is a specific result of our interaction model and of the overall scheme that we have set up. 

We now address the first two points that have been mentioned above. As shown by Eq.~(\ref{StateUnitaryT}), when the evolution of the system is purely unitary, the geometric phase picked up by the control qubit  does not depend on the amplitude $\alpha$ of the oscillator's initial state. This leaves the value of $\theta$ unchanged even if the initial state of the oscillator is a thermal state $\rho_V = \int d \alpha P(\alpha,0,V) | \alpha \rangle \langle \alpha | $, where $P(\alpha,\alpha_0,V)=\frac{2}{\pi(V-1)}e^{-\frac{2|\alpha-\alpha_0|^2}{V-1}}$ is the Gaussian thermal distribution centered at point $\alpha_0$ in the phase space and having variance $V = (e^\beta{+}1)/(e^\beta{-}1)$, with $\beta{=}\hbar\omega_m/k_b\mathcal{T}$ ($k_b$ is the Boltzmann constant and ${\cal T}$ is the temperature of the oscillator). By taking the initial state $\rho_0 = |+\rangle \langle + | \otimes \rho_V$, it is straightforward to see that a phase identical to the pure-state case is acquired by the qubit. In fact, the state at time $\tau$ reads $\rho(\tau) = |\varphi\rangle \langle \varphi| \otimes \rho_V$, where $|\varphi\rangle =(|1\rangle + e^{-i 2 \eta^2 \tau^2} |0\rangle)/\sqrt2$. The invariance of $\theta$ in such a mixed-state scenario has also been confirmed using the framework for the evaluation of mixed-state geometric phases proposed in Ref.~\cite{Eric}. This shows that ignorance on the initial preparation of the state of the harmonic oscillator does not affect the possibility to generate and detect a geometric phase. 

We now assess the potential effects that non-unitary dynamics may have on the occurrence of the geometric phase under scrutiny. On one hand, the consideration of an explicitly open dynamics will make our proposal closer to the reality of the potential experimental situations that will be addressed later on in this work. On the other hand, it is reasonable to expect significant deviations from the results found so far when we are far from unitarity. We thus consider the oscillator as affected by dissipation at rate $\gamma$ with the control qubit still evolving unitarily, a situation that is formally described by the master equation $\dot{\rho} = - (i/\hbar)[\hat H, \rho] + \hat{\mathcal{L}} \rho$, where $\hat{\mathcal{L}} \rho{=}\gamma(\hat b\rho \hat b^\dag{-}\{\hat b^\dag \hat b, \rho\}/2)$ formally describes the oscillator's damping. Although we do not refer to any explicit experimental configuration, at this stage, the analysis that we perform here adheres very well to the experimental observations on the open-system dynamics of mechanical systems availbale to date~\cite{glance,glance2}.  In order to grasp the effects of the dissipative dynamics, we divide the time-window of the evolution in small intervals, each of length $\delta t$. Inspired by the Suzuki-Trotter formula~\cite{trotter}, the dynamics can then be approximated by alternating a unitary evolution described by $\hat{\mathcal{U}}^{\varphi}_{\delta t}  \rho{=}\hat U_{\varphi}(\delta t) \rho \hat U_{ \varphi}^\dag(\delta t)$ and the purely dissipative propagator $\hat{\mathcal{D}}_{\delta t} \rho{=}e^{{\hat{\mathcal{L}}  \delta t}} \rho$~\cite{Jacob,commento}. After $N$ iterations, we have the evolved state
\begin{equation}
\rho(N \delta t){=} \big(\hat{\mathcal{D}}_{\delta t} \hat{\mathcal{U}}^{\varphi}_{\delta t}\big)^N \rho_0.
\end{equation}
This approach is particularly useful in analyzing a damped harmonic oscillator. Indeed, the action of the dissipative superoperator $\hat{\mathcal{D}}_t$ on the off-diagonal elements of a density matrix written in a coherent state basis is given by the dyadic expression~\cite{phoenix} 
\begin{equation}
\label{ph}
\hat{\mathcal{D}}_t |\lambda_1\rangle \langle \lambda_2| = \langle \lambda_2 | \lambda_1\rangle^{1 - \exp(-\gamma t)} | \lambda_1 e^{-\gamma t} \rangle \langle \lambda_2 e^{-\gamma t}|,
\end{equation}
where $|{\lambda_j}\rangle$ ($j=1,2$) are two coherent states. As we are interested in short-time intervals $\delta t$, we take $1 - \exp(-\gamma \delta t)\simeq\gamma \delta t$. Therefore, the action of $\hat{\mathcal{D}}_{\delta t}$ on the state of our system results in the displacement of the harmonic oscillator and the exponential decrease of its initial amplitude $\alpha$. Moreover, from Eq.~(\ref{ph}) we see that a phase factor is attached to the off-diagonal elements of the density matrix. Such features are useful to close the path across which the oscillator is displaced, as we now show using an example close, in spirit, to the previous one. Differently from the case of a unitary evolution, when dissipation is included the phase $\varphi$ is set to $0$ for $t \in [t_0, t_0 + T_1]$ and to $\pi$ for $t \in [t_0 + T_1,t_0 + T_1 +T_2]$.  The state at time $t_0+T_1$ is given by $\rho(t_0 +T_1) = (\hat{\mathcal{D}}_{\delta t} \hat{\mathcal{U}}_{\delta t}^{0})^{N_1} \rho_0$, with $N_1 \delta t = T_1$. Taking the limit $\delta t \rightarrow 0$, this turns out  to be
\begin{equation}
		\begin{aligned}
\rho(t_0+T_1)=&\frac{1}{2}\left( |1,\alpha_1 \rangle \langle 1,\alpha_1 | +  |0,\alpha_1  -  i \beta_1 \rangle \langle 0,\alpha_1  -  i \beta_1 |\right.\\
&+e^{-i \theta_1(\alpha)} e^{-\Gamma_1} |0,\alpha_1  - i \beta_1 \rangle \langle 1, \alpha_1 |+h.c. ),
		 \end{aligned}\label{T1}
	\end{equation}
where
\begin{equation}
		\begin{split}
\theta_1(\alpha) = &\frac{\eta\alpha}{2\gamma}  (1-e^{-2\gamma  T_1}),~~\beta_1=\frac{\eta}{\gamma} (1- e^{-\gamma T_1}),~~\alpha_1 = \alpha e^{-\gamma T_1},\\
&\Gamma_1 = \frac{\eta^2}{2\gamma^2} [\gamma T_1 +\frac{1}{2}(1-e^{-2\gamma T_1}) - 2(1 - e^{-\gamma T_1})].
	 	\end{split}
	\end{equation}
Eq.~(\ref{T1}) shows that the state of the oscillator at time $T_1$ is conditionally displaced by a quantity $-i\beta_1$. Moreover, the state of the qubit-oscillator system, which is damped at a rate $\Gamma_1$, aquires a phase factor $\theta(\alpha)$ that, differently from the unitary case, depends linearly on $\alpha$. We can then proceed to evauate the state of the system at time $t_0+T_1+T_2$ by setting $\varphi{=}\pi$ and taking $\rho(t_0+T_1+T_2) = (\hat{\mathcal{D}}_{\delta t} \hat{\mathcal{U}}_{\delta t}^{\pi})^{N_2} \rho(t_0+T_1)$, with $N_2 \delta t = T_2$. This displaces the state of the oscillator by $i \beta_2$ in the opposite direction to what occurred at $T_1$.  The time interval $T_2$ is chosen such that the oscillator displacement $-i \beta_1$ accumulated  during $T_1$ is cancelled~\cite{commentotime}. By calling $\tilde\tau =  T_1 +T_2$, the final state reads
\begin{equation}
		\begin{split}
\rho(t_0 + \tilde\tau)&=\frac{1}{2}\left[\openone  +e^{-\Gamma } (e^{-i \theta(\alpha)}  |0 \rangle \langle 1 |+h.c.) \right] \otimes |\alpha e^{-\gamma \tilde\tau}\rangle \langle \alpha e^{-\gamma \tilde\tau}|,
		 \end{split}\label{tau2}
	\end{equation}
where
$\theta(\alpha){=}{\eta\alpha}\frac{1-2 e^{-\gamma T_1} + e^{-2 \gamma T_1}}{\gamma(2-e^{-\gamma T_1})}$, $\Gamma{=}{\eta^2}\tilde{\Gamma}(\gamma,T_1,T_2)/{2\gamma^2}$
with $\tilde{\Gamma}(\gamma,T_1,T_2)$ a dimensionless function that behaves as $\gamma^3$ for $\gamma\rightarrow 0$, thus ensuring that $\Gamma \rightarrow 0$ as $\gamma \rightarrow 0$~\cite{Gamma}. It is easily seen that the phase $\theta(\alpha)$ gained by the oscillator in this process is equal to the area ${\cal A}$ enclosed by the displacement path in parameter space, as shown in Fig.~\ref{geomphase} {\bf (b)}. The detectability of such phase depends on the function $\Gamma$, which determine the decoherence rate for the off-diagonal terms in $\rho(t_0+\tilde\tau)$. Indeed, in order to achieve a non-vanishing  phase, we should ensure that $\Gamma{\ll}1$. Remarkably, while $\theta(\alpha)$ depends on the amplitude of the initial coherent state, $\Gamma$ does not. Surprisingly, by choosing $\eta/\gamma \ll 1$ (which embodies the weak-coupling condition between the oscillator and the control qubit), the requirement $\Gamma \ll 1$ is fulfilled. On the other hand, we can achieve any value of $\theta(\alpha)$ making an appropriated choice for the value of $\alpha$, so as to compensate the conditions required for a negligible damping.

\begin{figure}[t]
{\bf (a)}\hskip3.8cm{\bf (b)}
\includegraphics[width=0.49\columnwidth]{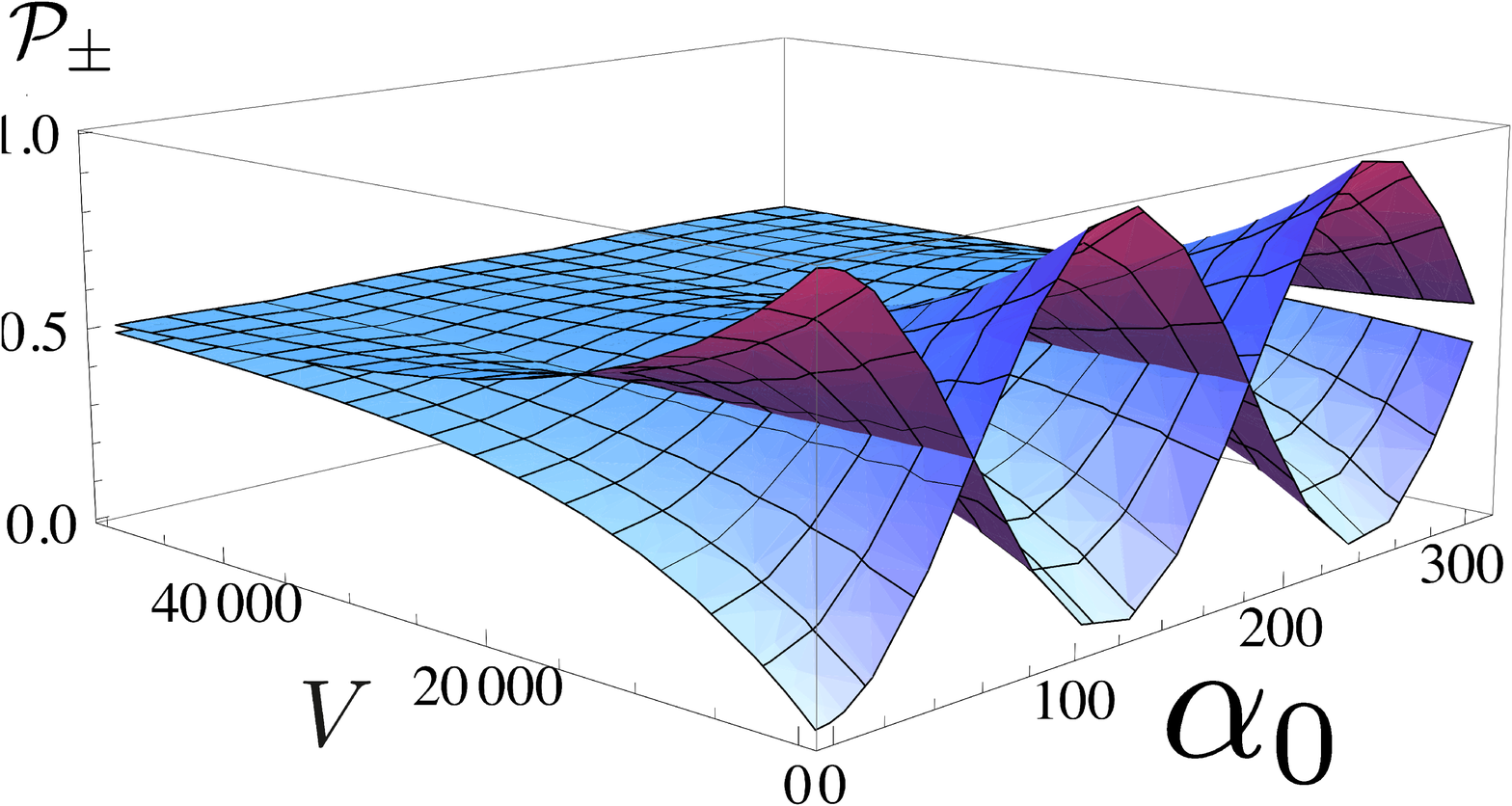}~
\includegraphics[width=0.49\columnwidth]{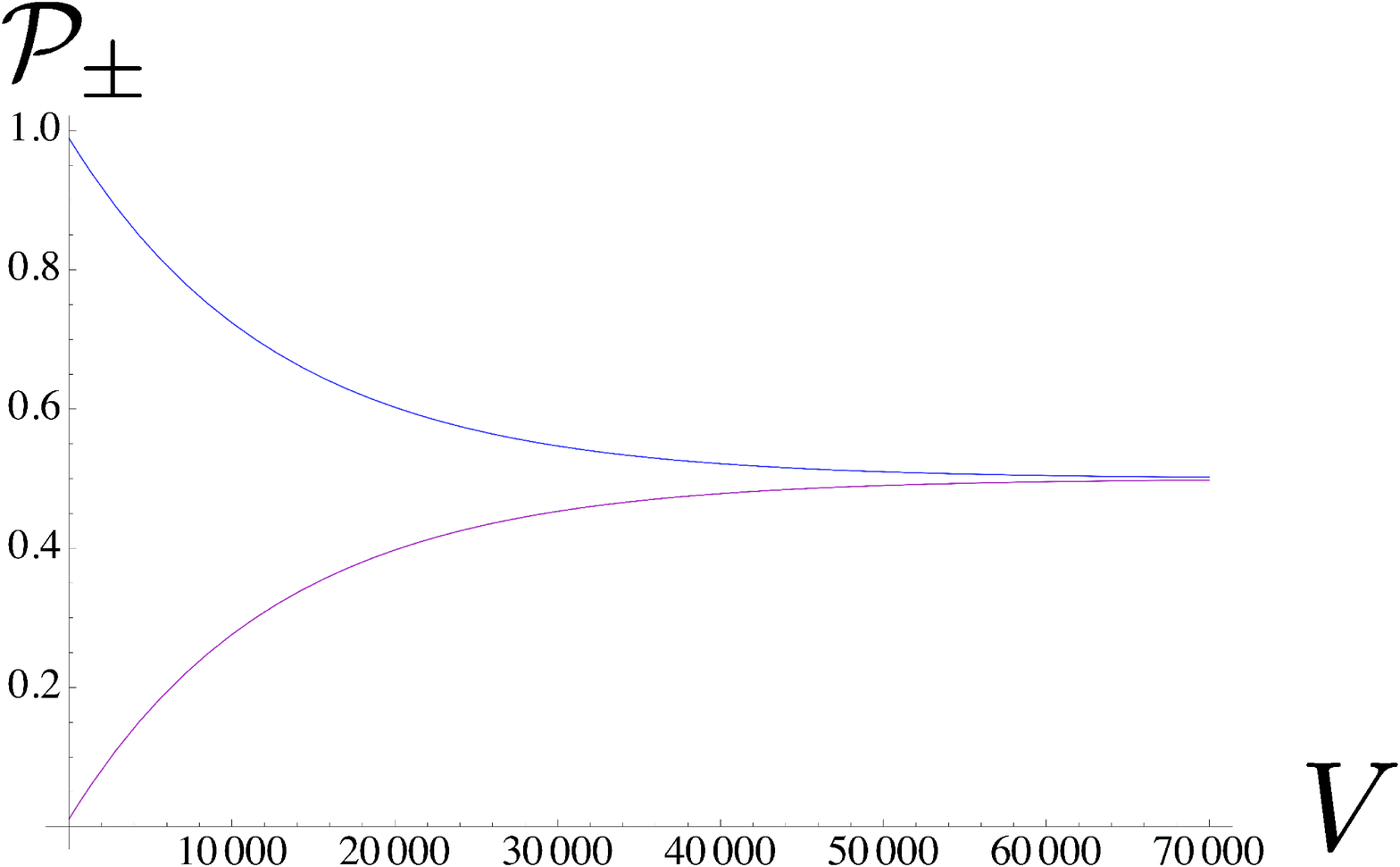}
\caption{{\bf (a)} Probabilities $\mathcal{P}_+$ and $\mathcal{P}_-$ against the displacement $\alpha_0$ and the parameter V for $\eta/\gamma = 0.05$ and $\gamma T_1= 20$. {\bf (b)} Same probabilities against $V$ for and $\alpha_0 = 0$ and the same parameters as in panel {\bf (a)}.}\label{Phase}
\end{figure}

The approach described above can be applied so as to evaluate the effects, on the geometric phase, due to a thermal preparation of the state of the oscillator undergoing dissipative dynamics. We thus assume that the initial state of the oscillator is the displaced thermal state $\rho^{\alpha_0}_V = \int d \alpha P(\alpha,\alpha_0,V) | \alpha \rangle \langle \alpha |$. Following the lines sketched so far, we arrive at the evolved state
\begin{equation}
\rho^{\alpha_0}_V(t_0 + \tilde\tau) = \int d^2\alpha P(\alpha,\alpha_0,V) \rho(t_0+\tilde\tau),\label{rhoV}
\end{equation}
where $ \rho(t_0+\tilde\tau)$ is given by Eq.~(\ref{tau2}). In light of the dependence of the phase $\theta$ on the amplitude $\alpha$ [as shown right after Eq.~(\ref{tau2})], the control qubit and the oscillator end up in a correlated state. This complicates the calculation of the overall geometric phase associated with $\rho^{\alpha_0}_V(t_0 + \tilde\tau)$. Nevertheless, it is still possible to evaluate the geometric phase by adopting the framework developed in Ref.~\cite{Eric}, which is based on the probability that a measurement over the state of the control qubit has outcomes $\{|+\rangle, |-\rangle\}$ (with $\ket{-}=(\ket{0}-\ket{1})/\sqrt2$). In order to understand this, let us  first consider the state in Eq. (\ref{tau2}) and suppose to project the control qubit onto the $\{\ket\pm\}$ basis. The corresponding outcome probabilities are given by $\mathcal{P}_\pm = \langle \pm |\text{Tr}_{m}\{ \rho(t_0+\tilde\tau)\} | \pm \rangle$, where $\text{Tr}_m$ denotes the partial trace over the oscillator's degrees of freedom. A straightforward calculation shows that  $\mathcal{P}_{\pm} =[1 \pm v \cos\theta(\alpha)]/2$,  where $v = e^{-\Gamma}$. This reminds us of the fringes of an interferometer whose visibility is $v$: the state of the composite qubit-oscillator system evolves along two branches (one associated to $\ket{1}$ and the other to $\ket0$) that can be seen as two arms of a Mach-Zehnder interferometer. The two components of the state of the system that have undergone the evolution ruled by $\hat H$ (and possibly the dissipative dynamics considered here) are then let interfere by projecting the qubit state onto the superposed basis. This analysis offers us an operative interpretation of $\theta(\alpha)$. In fact, by changing it, the probabilities $\mathcal{P}_{\pm}$ change, reaching the complete inversion when $\theta(\alpha) = \pi$. Therefore, $\theta(\alpha)$ can be operatively seen as the inversion of the outcome probabilities $\mathcal{P}_{\pm}.$  We can attach an analogous meaning to the phase associated with a mixed initial state. We thus now consider the state in Eq.~(\ref{rhoV}), which gives us $\mathcal{P}_{\pm} =(1 \pm |\Lambda| \cos[\tilde{\theta}])/2$ with $\tilde\theta=\arg({\Lambda})$ and

\begin{equation}
\Lambda=v \int d^2\alpha  P(\alpha,\alpha_0,V) e^{i \theta (\alpha)}.
\end{equation} 

As in the case of a pure state, the phase $\tilde{\theta}$ is operatively defined through the inversion of the probabilities ${\cal P}_\pm$. Fig.~\ref{Phase} {\bf (a)} shows such quantities against the initial displacement $\alpha_0$ and the temperature $V.$ Although the visibility of the fringes decreases with the increasing temperature, it is possible to see a population inversion even for high values of $V.$  Another very interesting situation is the one in which  the initial state is a non-displaced thermal state, i.e.  the Gaussian distribution in Eq. (\ref{rhoV}) is centered in $\zeta = 0.$ The behavior of the outcome probabilities $\mathcal{P}_\pm$ against the thermal variance $V$ is shown in Fig.~\ref{Phase} {\bf (b)}. A larger temperature results in an increase (decrease) of $\mathcal{P}_-$ ($\mathcal{P}_+$). The partial inversion of the probabilities is due to the average geometric phase $\tilde{\theta}(\alpha)$ picked up by the oscillator during the process.

In summary, we have shown how to generate a geometric phase on a system in which a qubit is coupled to a harmonic oscillator. The phase can be detected using the qubit as an interferometer. We propose systems combining effective two-level devices to mechanical modes as potential scenarios for the implementation of our proposal~\cite{glance,nanooscillator}. In the nano-scale domain, Hamiltonian models of a form close to the one proposed here can be achieved by capacitively combining a nano-cantilever to a Cooper-pair box or growing a quantum dot on a nano-beam~\cite{nanooscillator,nano}. At the microscopic scale, on the other hand, the coupling in Eq.~\eqref{Heff} can be engineered by means of a three-level atom trapped within the volume of a pumped optomechanical cavity field and off-resonantly coupled to the latter~\cite{noi}. Analogous configurations have been recently proposed~\cite{wallquist} as valid alternative to consolidated schemes for the coupling between a mechanical mode and the vibrational degrees of freedom of a single atom, an ensemble of them or a levitating nanoparticle~\cite{wallquist,oriol}. Under the presence of a dissipative environment and for a mixed thermal state of the oscillator, the geometric phase can still be observed under conditions over the coupling between the qubit and the oscillator that can be matched experimentally~\cite{glance,glance2}. For a nano-beam with fundamental frequency $\sim{100}$MHz coupled to a superconducting qubit at a rate $\eta\sim1$MHz and having a (realistic) decay rate of $\sim1-10$MHz, which are values well within the validity of our approach, a temperature of $0.5$K keeps the probabilities ${\cal P}_\pm$ at the visible level. By driving the mechanical mode with a two-tone signal~\cite{cohadon}, which is possible optically and electrically, thus covering both the micro- and nano-scale configurations. In the first scenario, one could consider, for instance, a single Cs atom coupled to a light mechanical resonator (masses are typically in the ng range) in both the end-mirror or membrane-in-the-middle arrangement~\cite{glance}, as recently considered for the problem of coupling the external degrees of freedom of an atom to the vibrations of a massive mechanical oscillator~\cite{wallquist}. High-finesse cavities with small waists are currently employed in controllable optomechanical experiments (finesse${\sim}10^5$ with a waist of a few $\mu$m), thus guaranteeing a strong enough light-atom interaction that is suitable for the achievement of the effective Hamiltonian model proposed in Eq.~\eqref{Heff}. For the examples discussed here, all the experimental observations that are currently available are in full agreement with a Markovian description of the dynamics induced by the thermal background of phonons affecting the mechanical oscillator, thus making the our study perfectly appropriate. Our proposal moves along the lines of an investigation assessing quantum effects in macroscopic systems. It enlarges the fan of indicators of quantumness at the meso-scale with a figure of merit, the geometric phase, that arises in virtue of the sole coherent qubit-oscillator interaction and survives against plainly adverse operating conditions.


MP and GMP thank the CQT, National University of Singapore, where part of this work has been done. We acknowledge financial support from the National Research Foundation and Ministry of Education in Singapore, the UK EPSRC, EUROTECH, EU-projects GEOMDISS, QNEMS and SOLID, the Royal Society and the Wolfson Trust. VV is a fellow of Wolfson College, Oxford.


\begin{thebibliography}{}

\bibitem{glance} M. Aspelmeyer, S. Groeblacher, K. Hammerer, and N. Kiesel, J. Opt. Soc. Am. B {\bf 27}, A189 (2010); T. J. Kippenberg and K. J. Vahala, Science {\bf 321}, 1172 (2008); F. Marquardt and S. M. Girvin, Physics {\bf 2}, 40 (1993).


\bibitem{mancini} S. Mancini, V. Giovannetti, D. Vitali, P. Tombesi, \prl {\bf 88}, 120401 (2006). 

\bibitem{mauro2}M. Paternostro, {\it et al.}, 
\prl  {\bf 99}, 250401 (2007).

\bibitem{marshall}W. Marshall, C. Simon, R. Penrose, D. Bouwmeester, \prl {\bf 91}, 130401 (2003). 

\bibitem{vlatko}A. Ferreira, A. Guerreiro, V. Vedral, \prl {\bf 96}, 060407 (2006). 

\bibitem{pirandola}S. Pirandola, D. Vitali, P. Tombesi, S. Lloyd, \prl {\bf 97}, 150403 (2006).

\bibitem{vitali} D. Vitali, {\it et al.},  
\prl {\bf 98}, 030405 (2007). 

\bibitem{glance2}  K. C. Schwab and M. L. Roukes, Phys. Today {\bf 58}, 36 (2005); M. Poot, H. S. J. van der Zant, arXiv:1106.2060 (to appear in Phys. Report, 2011).

\bibitem{nanooscillator} P. Rabl, A. Shnirman, and P. Zoller, Phys. Rev. B {\bf 70}, 205304 (2004); M. D. LaHaye, O. Buu, B. Camarota, and K. C. Schwab, Science {\bf 304}, 74 (2004); A. Naik, {\it et al.}, Nature (London) {\bf 443}, 193 (2006); S. Pugnetti, Y.M. Blanter, and R. Fazio, Europhys. Lett. {\bf 90}, 48007 (2010).

\bibitem{noi} G. Vacanti, {\it et al.} (submitted, 2010).

\bibitem{berry} M. V. Berry, Proc. Roy. Soc. A, {\bf 329}, 45 (1984).

\bibitem{aharonov} Y. Aharonov, J. Anandan, \prl {\bf 58}, 1593 (1987).

\bibitem{angelogp1} A. Carollo, I. Fuentes-Guridi, M. Franca Santos, V. Vedral, \prl {\bf 90}, 160402 (2003); \prl {\bf 92}, 020402 (2004).

\bibitem{wineland} D. Leibfried, {\it et al.}, Nature (London) {\bf 422},  412 (2003).

\bibitem{milburn} G. J. Milburn, S. Scheneider, and D. F. V. James, Fortschr. Phys.  {\bf 48},  801 (2000).

\bibitem{nota} $\Im\left( \oint \zeta^*d\zeta \right) = \oint xdy - ydx = \int_{\sigma} dxdy = \mathcal{A}$, where we have taken $\zeta = x + iy$.

\bibitem{notabis} The same result, {\it i.e.} an overall phase $\vartheta_{\text{tot}}$ equal to $\mathcal{A}$ for an arbitrary closed loop is obtained using Anandan's rule $\vartheta _{\text{tot}}{=}\theta_D+\theta_G$ with $\theta_D{=}-1/\hbar \int_{0}^{\tau} \langle \psi(t) | H(t) |\psi(t) \rangle dt $ the dynamical phase, $\theta_G{=}i \int_{0}^{\tau} \langle \psi(t) | \partial_t |\psi(t) \rangle dt$ the geometric one~\cite{aharonov} and $\tau$, $\tau_0$ the initial and final time of the evolution.  

\bibitem{Eric} E. Sjoqvist, {\it et al.}, 
Phys. Rev. Lett. {\bf 85}, 2845 (2000).

\bibitem{trotter} H. F. Trotter, Proc. Am. Math. Soc. {\bf 10}, 545 (1959); M. Suzuki, Commun. Math. Phys. {\bf 51}, 183 (1976).

\bibitem{commento} For the Hamiltonian consider here, the chosen approach is an excellent approximation of the exact dynamics of the system. Introducing the superoperator $\hat{\mathcal{H}} \rho = - i[ \hat H, \rho]$, we write the formal solution of the open-system dynamics as $\rho(t) = e^{ \hat{\mathcal{H}} t  + \hat{\mathcal{L}} t } \rho_0.$ Upon explicit calculation, it is straightforward to show that, as $\delta t \rightarrow 0$, $[ \hat{\mathcal{D}}_{\delta  t}, \hat{\mathcal{U}}^{\varphi}_{\delta  t}]=0$   
so that $\rho(N\delta{t})\rightarrow \rho( t) $ in this limit. 

\bibitem{Jacob} H. Jeong, Phys. Rev. A {\bf 72}, 034305 (2005).

\bibitem{phoenix} S. J. D. Phoenix, \pra {\bf 41}, 5132 (1989).

\bibitem{commentotime} This condition can be met by taking $T_{1,2}$ such that $e^{-\gamma T_1}{=}2{-}e^{\gamma T_2}$. 

\bibitem{Gamma} The complete expression of $\tilde{\Gamma}(\gamma,T_1,T_2)$ is $\tilde{\Gamma}(\gamma,T_1,T_2){ =} \gamma \tau{+}\frac{1}{4} (1{-}e^{-\gamma T_1})^2 (1{-}e^{-2 \gamma T_2}){+}\frac{1}{2} (2{-}e^{-2\gamma T_1}{-}e^{-2 \gamma T_2})-2(2{-}e^{-\gamma T_1}{-}e^{-\gamma T_2}){-}(1-e^{-\gamma T_1}) [ (1-e^{-\gamma T_2})-\frac{1}{2} (1-e^{-2\gamma T_2})].$ Truncating the expansion of the exponential function to the second order in $\gamma T_j{\ll}{1}$, the only non-null terms are those $\propto\gamma^3.$  

\bibitem{nano} I. Wilson-Rae, P. Zoller, and A. Imamoglu, Phys. Rev. Lett. {\bf 92}, 075507 (2004).

\bibitem{wallquist} M. Wallquist, {\it et al.}, Phys. Rev. A {\bf 81}, 023816 (2010); K. Hammerer, {\it et al.}, Phys. Ref. Lett. {\bf 103}, 063005 (2009). 

\bibitem{oriol} D. E. Chang, {\it et al.}, Proc. Nat. Acad. Sci. USA {\bf 107}, 1005 (2010); O. Romero-Isart, {\it et al.}, New J. Phys. {\bf 12}, 033015 (2010); M. Paternostro, G. De Chiara, and G. M. Palma, Phys. Rev. Lett. {\bf 104}, 243602 (2010); C. Genes, D. Vitali, and P. Tombesi, Phys. Rev. A {\bf 77}, 050307(R) (2008); G. De Chiara, M. Paternostro, and G. M. Palma, {\it ibid.} {\bf 83}, 052324 (2011).

\bibitem{cohadon} P. Verlot, {\it et al.}, Phys. Rev. Lett. {\bf 104}, 133602 (2010).






\end{thebibliography}
\end{document}